\documentclass[aps,prl,twocolumn,showpacs,amsmath,amssymb]{revtex4}
\usepackage{graphicx}
\usepackage{dcolumn}
\usepackage{bm}

\begin{document}

\title{A standard Hamiltonian formulation for the dynamical Casimir effect}

\author{Toru Kawakubo and Katsuji Yamamoto}
\affiliation{Department of Nuclear Engineering, Kyoto University,
Kyoto 606-8501, Japan}
\date{\today}

\begin{abstract}
We present a quantum description of photon creation
via dynamical Casimir effect based on the standard Hamiltonian formulation.
The particle representation is constructed
in the expansion of field operators fixed with the initial modes.
The Hamiltonian is presented
in terms of the creation and annihilation operators
with the time-varying couplings
which originate from the external properties
such as an oscillating boundary or a plasma mirror of a semiconductor slab.
Some consideration is also made for the experimental realization
with a semiconductor plasma mirror.
\end{abstract}


\maketitle

{\it Introduction}.--The quantum nature of vacuum provides
a variety of physically interesting phenomena, including the Casimir effect
\cite{C48}.
The so-called dynamical (non-stationary) Casimir effect (DCE),
as well as the static force, has been investigated extensively
\cite{P68-69,M70,FD76,RT85,JS96,BE93-BC95-CB95,LJR96,
Law94,SPS98,D95-DK96,D98,CDM01-02,SH02,IT04-05,
Y89,LTV95,CDLM04,UPSS04,DD05-06,R06,HE06-07} (also references therein),
where photons are created from the vacuum fluctuation
in non-adiabatic change of the system
driven by vibration of a cavity or expansion of the universe.
Most of the theoretical approaches are based on the field expansion
in terms of the instantaneous modes.
Since there is no unitary map among the instantaneous modes
with different boundary conditions,
the Hamiltonian formulation to describe the time-evolution
does not exist in the standard sense
\cite{M70}.
Even in this case, the time-evolution of the instantaneous-mode operators
may be described by means of an effective Hamiltonian
\cite{Law94,SPS98}.
On the other hand, a suitable transformation
may be made to move to a specific coordinate system
with fixed boundaries for canonical quantization
\cite{RT85,JS96,HE06-07,BE93-BC95-CB95}.

Experimentally, it is difficult to realize
a sufficient magnitude of mechanical vibration
at a resonant frequency $ \sim $ GHz
to create a significant number of photons for detection.
As a feasible alternative, it has been proposed recently
that the oscillating wall can be simulated
by a plasma mirror of a semiconductor slab
irradiated by periodic laser pulses \cite{IT04-05}.
(See also Refs. \cite{Y89,LTV95}.)

In this paper, we investigate a quantum description of DCE,
presenting the standard Hamiltonian to govern the unitary time-evolution.
We are particularly concerned with
the experimental realization of DCE with a semiconductor plasma mirror.
It is indeed important to develop the Hamiltonian formulation
to investigate the quantum properties of the system,
including the detection of created photons
through interaction with suitable probe such as atoms.
This standard description has several advantages:
(1) The particle representation is constructed
in the expansion of field operators fixed with the initial modes.
It is neither necessary to trace the mode change in time,
nor to seek a specific coordinate system for quantization.
Simple formulas are presented to calculate the couplings
among the creation and annihilation operators in the Hamiltonian
as space-integrals involving the initial mode functions.
Similar formulas are obtained
for the instantaneous-mode effective Hamiltonian
where the time-derivatives of the mode functions are further involved
\cite{Law94,SPS98}.
(2) The time-variation of the creation and annihilation operators
in the standard formulation precisely represents
the unitary quantum evolution all the time.
On the other hand, the time-variation
of the instantaneous-mode operators
and that of the mode functions together provide the quantum evolution.
The instantaneous-mode operators and mode functions
coincide with those of the standard formulation
just at each period of the oscillation.
(3) The present formulation is applicable
to various physical setups, including the oscillating wall
and the semiconductor plasma mirror, as investigated later.
We can check that under the situations
where the mode functions do not change largely in time,
as usually considered,
this standard description provides essentially the same result for DCE
as the instantaneous-mode description.
Furthermore, the Hamiltonian is readily calculated
even for the large time-variation of external properties,
clarifying the dependence on experimental parameters
specifically for the plasma mirror case.

{\it Standard Hamiltonian formulation}.--We consider
a scalar field generally in 3+1 space-time dimensions.
The Lagrangian is given as
\begin{equation}
{\cal L} = \frac{1}{2} \epsilon ({\bf x},t) ( {\dot \phi} )^2
- \frac{1}{2} ( \nabla \phi )^2
- \frac{1}{2} m^2 ({\bf x},t) \phi^2
\end{equation}
($ \hbar = c = 1 $)
\cite{Law94,SPS98,UPSS04,CDLM04,BE93-BC95-CB95}.
Here, $ \epsilon ({\bf x},t) $ and $ m^2 ({\bf x},t) $
represent the dielectric permittivity
and conductivity (effective ``mass" term), respectively,
in the matter region such as a semiconductor slab.
As specified later, they are space-time dependent,
simulating the boundary oscillation.
Conventionally, the instantaneous modes $ {\bar f}_\alpha ({\bf x},t) $
(real, orthonormal and complete) at each time $ t $
with time-varying frequencies $ {\bar \omega}_\alpha (t) $
are adopted according to the boundary oscillation:
$ [ - \nabla^2 + m^2 ({\bf x},t) ] {\bar f}_\alpha ({\bf x},t)
= \epsilon ({\bf x},t) {\bar \omega}_\alpha^2 (t)
{\bar f}_\alpha ({\bf x},t) $
with $ \int_V \epsilon ({\bf x},t) 
{\bar f}_\alpha ({\bf x},t) {\bar f}_\beta ({\bf x},t) d^3 x
= \delta_{\alpha \beta} / [ 2 {\bar \omega}_\alpha (t) ] $.
Instead, we here construct the particle representation
in terms of the initial modes
\begin{eqnarray}
f^0_\alpha ({\rm x}) = {\bar f}_\alpha ({\bf x},t=0) , \
\omega^0_\alpha = {\bar \omega}_\alpha (t=0) .
\end{eqnarray}
The canonical field operators in the Heisenberg picture
are expanded with the creation and annihilation operators
$ a_\alpha^\dagger (t) $ and $ a_\alpha (t) $ as
\begin{eqnarray}
\phi ({\bf x},t)
&=& \sum_\alpha [ a_\alpha (t) + a_\alpha^\dagger (t) ]
f^0_\alpha ({\bf x}) ,
\\
\Pi ({\bf x},t)
&=& \epsilon ({\bf x},0)
\sum_\alpha i \omega^0_\alpha [ - a_\alpha (t) + a_\alpha^\dagger (t) ]
f^0_\alpha ({\bf x}) ,
\end{eqnarray}
where $ \Pi ({\bf x},t) = \partial {\cal L} / \partial {\dot \phi}
= \epsilon ({\bf x},t) {\dot \phi} ({\bf x},t) $.
Then, the Hamiltonian (Schr\"odinger picture) is presented
by the usual procedure as
\begin{eqnarray}
H(t) &=& \int_V \frac{1}{2} \left\{ \frac{\Pi^2}{\epsilon ({\bf x},t)}
+ \phi [ - \nabla^2 + m^2 ({\bf x},t) ] \phi \right\} d^3x
\nonumber \\
&=& \sum_{\alpha}
\omega_{\alpha} (t) \left( a_\alpha^\dagger a_\alpha + \frac{1}{2} \right)
+ \sum_{\alpha \not= \beta}
\mu_{\alpha \beta} (t) a_\alpha^\dagger a_\beta
\nonumber \\
&{}& + \sum_{\alpha , \beta}
i \left[ g_{\alpha \beta} (t) a_\alpha^\dagger a_\beta^\dagger
- g_{\alpha \beta}^* (t) a_\beta a_\alpha \right] ,
\label{eqn:Ht}
\end{eqnarray}
where the space-integral is taken over the whole region $ V $
which is ``fixed" (not time-dependent) according to the physical setup,
as explicitly shown later for typical cases.
The Heisenberg and Schr\"odinger pictures
are related by the unitary transformation
$ U(t) $ of time-evolution generated by this Hamiltonian $ H(t) $
as $ a_\alpha (t) = U^\dagger (t) a_\alpha U(t) $.
Hence, the time-dependence of the couplings in Eq. (\ref{eqn:Ht}),
which originates from the c-number external quantities
$ \epsilon ({\bf x},t) $ and $ m^2 ({\bf x},t) $,
is common in any pictures related by unitary transformations.
The Hamiltonian in the Heisenberg picture
$ H_{\rm H} (t) = U^\dagger (t) H(t) U(t) $
is obtained simply by $ a_\alpha \rightarrow a_\alpha (t) $,
$ a_\alpha^\dagger \rightarrow a_\alpha^\dagger (t) $,
$ \phi \rightarrow \phi ({\bf x},t) $,
$ \Pi \rightarrow \Pi ({\bf x},t) $ in $ H(t) $.

The mode frequencies $ \omega_{\alpha} (t) $,
intermode couplings $ \mu_{\alpha \beta} (t) $
and squeezing terms $ g_{\alpha \beta} (t) $ are calculated
by considering the orthonormality of $ f^0_\alpha ({\bf x}) $
which obey the wave equation
with $ \epsilon ({\bf x},0) $ and $ m^2 ({\bf x},0) $:
\begin{eqnarray}
\omega_{\alpha} (t) &=& \omega^0_\alpha + \mu_{\alpha \alpha} (t)
\equiv \omega^0_\alpha + \delta \omega_{\alpha} (t) ,
\label{eqn:omt}
\\
\mu_{\alpha \beta} (t)
&=& 2 G^\epsilon_{\alpha \beta} (t) + 2 G^m_{\alpha \beta} (t) ,
\label{eqn:mut}
\\
g_{\alpha \beta} (t)
&=& - i [ - G^\epsilon_{\alpha \beta} (t) + G^m_{\alpha \beta} (t) ] ,
\label{eqn:gt}
\\
G^\epsilon_{\alpha \beta} (t)
&=& \frac{1}{2} \omega^0_\alpha \omega^0_\beta \int_{\delta V (t)}
\frac{\epsilon^2 ({\bf x},0)}{\epsilon_\Delta ({\bf x},t)}
f^0_\alpha ({\bf x}) f^0_\beta ({\bf x}) d^3 x ,
\label{eqn:Ge}
\\
G^m_{\alpha \beta} (t)
&=& \frac{1}{2} \int_{\delta V (t)}
m_\Delta^2 ({\bf x},t) f^0_\alpha ({\bf x}) f^0_\beta ({\bf x}) d^3 x .
\label{eqn:Gm}
\end{eqnarray}
The integrals for $ G^{\epsilon,m}_{\alpha \beta} (t) $
are evaluated in practice in the subregion $ \delta V (t) $
($ \subseteq V $), possibly time-dependent,
where $ \epsilon ({\bf x},t) $ and $ m^2 ({\bf x},t) $ vary in time as
$ \epsilon_\Delta^{-1} ({\bf x},t)
\equiv \epsilon^{-1} ({\bf x},t) - \epsilon^{-1} ({\bf x},0) $
and $ m_\Delta^2 ({\bf x},t) \equiv m^2 ({\bf x},t) - m^2 ({\bf x},0) $.
[$ G^{\epsilon,m}_{\alpha \beta} (0) = 0 $ at $ t = 0 $
with $ H(0) $ diagonalized in terms of $ f^0_\alpha ({\bf x}) $.]
In order to demonstrate
the relevance of this Hamiltonian formulation for DCE,
we investigate two typical cases in the effective 1+1 dimensions
(1) oscillating wall and (2) plasma mirror of a semiconductor slab.

{\it Oscillating wall}.--The boundary walls may be represented
by high potential barriers of matter
extending infinitely (or finite and long) outside the cavity as
\begin{eqnarray}
m^2 (x,t) = m^2
[ - \infty < x < \delta (t) , L < x < \infty ] .
\label{eqn:m2-om}
\end{eqnarray}
Here, the right side is fixed at $ x = L $,
while the left side varies in time around $ x = 0 $
as $ \delta (0) = 0 \leq \delta (t) \leq \delta_1 \ll L $.
The dielectric is taken uniformly as $ \epsilon (x,t) = \epsilon_1 $
in the matter region.
This setup with potential barriers,
rather than the rigid boundary conditions,
may be similar to Ref. \cite{SH02}
where the matter-field interaction is treated dynamically.
The mode functions are given as
\begin{eqnarray}
{\bar f}_k (x,t) = \left\{ \begin{array}{ll}
C {\rm e}^{| k^\prime | ( x - \delta (t) )}
& ( - \infty , \delta (t) ):{\rm wall} \\
A \sin k [ x - \delta (t) + \xi ]
& [ \delta (t) , L ] \\
B {\rm e}^{- | k^\prime | ( x - L )}
& ( L , + \infty ):{\rm wall}
\end{array} \right.
\label{eqn:fkb-om}
\end{eqnarray}
with the dispersion relations
$ {\bar \omega}_k^2 = ( k^2 + k_\bot^2 ) / \epsilon_0
= ( {k^\prime}^2 + k_\bot^2 + m^2 ) / \epsilon_1 $,
where $ {k^\prime}^2 \simeq - m^2 < 0 $
and $ m \simeq | k^\prime | \gg k \sim 1/L $
for the large $ m^2 $,
and $ k_\bot $ ($ \sim k $) is the momentum
in the orthogonal spatial 2 dimensions
\cite{D98,CDM01-02,R06}.

For the large potential barrier $ m^2 $,
the frequency modulation $ \delta {\bar \omega}_k (t) $
and the diagonal squeezing coupling $ {\bar g}_{kk} (t) $
in the instantaneous-mode Hamiltonian
\cite{Law94,SPS98} are calculated by noting
$ \sin k [ L - \delta (t) + \xi ] \simeq 0 $ at $ x = L $
with $ \xi \simeq 1 / m \ll L $ and $ |B/A|, |C/A| \sim k \xi \ll 1 $
as
\begin{eqnarray}
\delta {\bar \omega}_k (t)
\simeq \omega^0_k [ \delta (t) / L ] r_k , \
{\bar g}_{kk} (t)
\simeq \delta {\dot{\bar \omega}}_k (t) / [ 4 {\bar \omega}_k (t) ] ,
\label{eqn:dombgb-om}
\end{eqnarray}
where $ r_k = k^2 / \epsilon_0 ( \omega^0_k )^2 $.
(The dielectric contribution is suppressed significantly
by $ \epsilon_1 {\bar \omega}_k^2 / m^2 \ll 1 $.)
This leading result is independent of the large $ m^2 $.
By taking formally the limit $ m \rightarrow \infty $
($ \xi \rightarrow 0 $), the usual moving boundary conditions
$ {\bar f}_k ( \delta (t) ,t ) = {\bar f}_k ( L , t ) = 0 $
are reproduced.

On the other hand, in the present standard description,
the frequency modulation and the squeezing term are calculated
in Eqs. (\ref{eqn:omt})--(\ref{eqn:Gm})
with $ \epsilon_0 ( \omega^0_k )^2 / m^2 \ll 1 $ as
\begin{eqnarray}
\delta \omega_k (t) & \simeq & 2i g_{kk} (t)
\simeq m^2 \int_0^{\delta (t)} A^2 \sin^2 k ( x + \xi ) dx
\nonumber \\
& \simeq & \left\{ \begin{array}{ll}
\delta {\bar \omega}_k (t) & [ m \delta (t) \ll 1 ] \\
\delta {\bar \omega}_k (t) [ m \delta (t) ]^2 / 3
& [ m \delta (t) \gg 1 ] \\
\end{array} \right. ,
\label{eqn:domg-om}
\end{eqnarray}
where $ f^0_k (x) = A \sin k ( x + \xi ) $ in $ 0 \leq x \leq L $
with $ \delta (0) = 0 $, $ \xi \simeq 1/m $,
$ A \simeq ( L \epsilon_0 \omega^0_k )^{-1/2} $ (normalization),
and $ m_\Delta^2 (x,t) = m^2 $ in $ \delta V (t) = ( 0 , \delta (x) ) $.
In the limit $ m \rightarrow \infty $
the standard $ \delta \omega_k (t) $ $ \propto [ m \delta (t) ]^2 $
diverges except at $ t = 0 $ with $ \delta (0) = 0 $,
while the instantaneous-mode $ \delta {\bar \omega}_k (t) $
remains finite as Eq. (\ref{eqn:dombgb-om}).
This corresponds to the claim that the Hamiltonian
does not exist in the moving boundary problem
\cite{M70}.
The squeezing couplings
$ {\bar g}_{kk} (t) $ in Eq. (\ref{eqn:dombgb-om})
and $ g_{kk} (t) $ in Eq. (\ref{eqn:domg-om}) appear to be different
even with $ \delta {\bar \omega}_k (t) \simeq \delta \omega_k (t) $.
It, however, will be shown that
they provide essentially the same result
for the photon creation at the resonance.

{\it Plasma mirror}.--We next consider the case of plasma mirror
which is realized with a semiconductor slab irradiated
by periodic laser pulses \cite{IT04-05}.
The dielectric response of plasma is given by
$ \epsilon ( \omega ) = \epsilon_1 [ 1 - ( \omega_p^2 / \omega^2 ) ] $
with the plasma frequency
$ \omega_p = ( n_e e^2 / \epsilon_1 m_* )^{1/2} $
in terms of the effective electron mass $ m_* $
and the conduction electron number density $ n_e $
proportional to the laser power $ W_{\rm laser} $.
The dispersion relation in plasma
$ k^2 = \epsilon ( \omega ) \omega^2
= \epsilon_1 \omega^2 - ( n_e e^2 / m_* ) $
can be taken into account in the slab region $ [ l , l + \delta ] $
around $ x = l $ with a thickness $ \delta ( \ll L ) $ as
\begin{eqnarray}
\epsilon (x,t) = \epsilon_1 (t) ,
m^2 (x,t) = m_p^2 (t) \equiv n_e (t) e^2 / m_* ,
\label{eqn:em-xt}
\end{eqnarray}
where $ m_p^2 (0) = 0 $ for $ W_{\rm laser} (0) = 0 $.
(The spatial distribution of the conduction electrons
may also be considered readily.)
The mode functions are given as
\begin{equation}
{\bar f}_k (x,t) = \left\{ \begin{array}{ll}
D \sin k x & [ 0 , l ) \\
B {\rm e}^{i k^\prime x} + C {\rm e}^{- i k^\prime x}
& [ l , l + \delta ]:{\rm slab} \\
A \sin k [ x - \delta + \xi (t) ]
& ( l + \delta , L ] \end{array} \right.
\label{eqn:fkb-pm}
\end{equation}
($ k^\prime = i | k^\prime | $ for $ {k^\prime}^2 < 0 $
with large $ m_p^2 $).
The Dirichlet boundary condition (corresponding to the TE mode)
is adopted at $ x = 0 , L $ with $ \sin k [ L - \delta + \xi (t) ] = 0 $.

The standard $ \delta \omega_k (t) $ and $ g_{kk} (t) $ are calculated
in Eqs. (\ref{eqn:omt})--(\ref{eqn:Gm})
with Eq. (\ref{eqn:fkb-pm}) for $ f^0_k (x) $ at $ t = 0 $ as
\begin{eqnarray}
\delta \omega_k (t)
&=& \omega^0_k [ \delta_\epsilon (t) + \delta_m (t) ] / L ,
\label{eqn:dom-pm}
\\
g_{kk} (t)
&=&
(i/2) \omega^0_k [ - \delta_\epsilon (t) + \delta_m (t) ] / L .
\label{eqn:g-pm}
\end{eqnarray}
Here, the effective wall oscillation is enhanced as
\begin{eqnarray}
\delta_\epsilon (t) / \delta & \simeq &
- [ \epsilon_1 (0) / \epsilon_0 ] [ 1 - \epsilon_1 (0) / \epsilon_1 (t)]
\sin^2 kl ,
\label{eqn:dlt-e}
\\
\delta_m (t) / \delta & \simeq &
[ m_p^2 (t) / \epsilon_0 ( \omega^0_k )^2 ]
\sin^2 kl .
\label{eqn:dlt-m}
\end{eqnarray}
This effect is almost proportional to
the square of mode function around the slab
$ [ f^0_k (l) ]^2 \propto \sin^2 kl $
since $ \int_l^{l+\delta} [ f^0_k (x) ]^2 dx
\simeq [ f^0_k (l) ]^2 \delta $
for $ k^\prime \delta
\sim [ \epsilon_1 (0) / \epsilon_0 ]^{1/2} ( \delta / L ) \ll 1 $
at $ t = 0 $.
If the slab is placed at the boundary $ x = l = 0 $,
$ \sin^2 kl $ is replaced
with $ ( k \delta )^2 / 3 \sim ( \delta / L )^2 \ll 1 $,
as observed in Ref. \cite{UPSS04}
claiming that DCE is suppressed in the TE mode.
The significant photon creation, however, will take place
even in the TE mode if the slab is placed
apart from the boundaries $ x = 0 , L $
which are the nodes of $ f^0_k (x) $
\cite{CDLM04,RITZ06}.

The shift $ \xi (t) $ in the instantaneous modes
of Eq. (\ref{eqn:fkb-pm}) is determined
mainly proportional to $ \delta $
to give the frequency modulation $ \delta {\bar \omega}_k (t) $.
The squeezing coupling $ {\bar g}_{kk} (t) $ is then calculated
with the formulas for the effective Hamiltonian
\cite{Law94,SPS98}.
After some calculations we find again
the relations $ \delta {\bar \omega}_k (t) \simeq \delta \omega_k (t) $
and $ {\bar g}_{kk} (t)
\simeq [ i / 2 {\bar \omega}_k (t) ] {\dot g}_{kk} (t) $,
as seen in Eqs. (\ref{eqn:dombgb-om}) and (\ref{eqn:domg-om})
for the oscillating wall,
where the change of dielectric is assumed to be small,
$ | \epsilon_1 (t) - \epsilon_1 (0) | \ll \epsilon_1 (0) $,
as usual \cite{UPSS04}.
This ensures the same result on DCE in both the descriptions
for the case of small oscillation, as shown later.

The above calculations of $ \delta \omega_k (t) $ and $ g_{kk} (t) $
are valid even for the large $ \epsilon_1 (t) $ and $ m^2_p (t) $
to provide the enhanced displacement
$ | \delta_{\epsilon, m} (t) | \gg \delta $,
which will be plausible experimentally.
It is not necessary here to consider
the large deformation of the mode functions in time
which invalidates the usual perturbative calculation
assuming the small change of the instantaneous modes.

{\it Photon creation as squeezing}.--Once the Hamiltonian
is presented in terms of the creation and annihilation operators,
the quantum properties of the system are investigated
readily by using the methods of quantum optics.
We here consider the quantum evolution for DCE,
restricted to a single resonant mode with time-varying frequency
$ \omega (t) = \omega_0 + \delta \omega (t) $
and squeezing coupling $ g(t) $, omitting the mode index ``$ k $".
The intermode couplings will not provide significant contributions
\cite{D95-DK96,CDM01-02,R06},
since generally due to the non-equidistant frequency differences
they are highly oscillating in the rotating-wave frame (interaction picture)
where the term $ \langle \omega \rangle a^\dagger a $ is eliminated
for the average frequency
$ \langle \omega \rangle = \omega_0 + \langle \delta \omega \rangle$
over the period $ T = 2 \pi / \Omega $ of the laser pulse.

The Heisenberg equation
$ i {\dot a} (t) = [ a(t) , H_{\rm H} (t) ] $
is described as the master equation,
\begin{eqnarray}
{\dot A} = - i \omega (t) A + 2 g (t) B , \
{\dot B} = i \omega (t) B + 2 g^* (t) A ,
\label{eqn:master}
\end{eqnarray}
in terms of the Bogoliubov transformation,
\begin{eqnarray}
a(t) = A(t) a + B^* (t) a^\dagger ,
a^\dagger (t) = A^* (t) a^\dagger + B (t) a .
\end{eqnarray}
The solution is expressed as
$ A(t) = \cosh r(t) {\rm e}^{i \phi_A (t)} $,
$ B(t) = \sinh r(t) {\rm e}^{i \phi_B (t)} $,
ensuring $ | A(t) |^2 - | B(t) |^2 = 1 $
with $ A(0) = 1 , B(0) = 0 $.
The unitary time-evolution is then given as a phase rotation and squeezing,
\begin{eqnarray}
U(t) = {\rm e}^{i K(t)}
{\rm e}^{- [ \lambda^* (t) a a - \lambda (t) a^\dagger a^\dagger ] /2 }
{\rm e}^{i \phi_A (t) a^\dagger a}
\end{eqnarray}
with $ \lambda (t) = r(t) {\rm e}^{i [ \phi_A (t) - \phi_B (t) ]} $
\cite{P68-69}.
The phase factor $ {\rm e}^{i K(t)} $
with $ K(t) = \phi_A (t) + \int_0^t \omega ( t^\prime ) / 2 d t^\prime $
is included to reproduce the zero-point energy of $ H(t) $
in $ i {\dot U} (t) = H(t) U(t) $.

An analytic solution for $ A(t) $ and $ B(t) $
is obtained in the rotating-wave approximation
by replacing $ \omega (t) \rightarrow
\omega_0 + \langle \delta \omega \rangle $ (average),
$ g (t) \rightarrow \langle g \rangle_\Omega {\rm e}^{-i \Omega t} $
(Fourier component).
By noting the time-evolution of the number operator
$ a^\dagger (t) a(t) = | B(t) |^2 a a^\dagger + \ldots $,
we obtain the photon creation via DCE (vacuum squeezing) as
\begin{eqnarray}
N_\gamma (t) = \langle 0 | a^\dagger (t) a(t) | 0 \rangle
\simeq (| 2  \langle g \rangle_\Omega | / \chi )^2
\sinh^2 \chi t
\label{eqn:Ngamma}
\end{eqnarray}
with the effective squeezing
\begin{eqnarray}
\chi = {\sqrt{| 2 \langle g \rangle_\Omega |^2 - \Delta^2}} ,
\label{eqn:chi}
\end{eqnarray}
allowing for the detuning $ \Delta $ of the laser pulse
\cite{D98,CDM01-02} as
\begin{eqnarray}
\Omega = 2 ( \omega_0 + \langle \delta \omega \rangle + \Delta ) .
\end{eqnarray}
The resonance for DCE is given precisely
by $ \Omega = 2 ( \omega_0 + \langle \delta \omega \rangle ) $
rather than $ \Omega = 2 \omega_0 $,
as considered in the instantaneous-mode approach
\cite{CDLM04}.
If $ \Omega = 2 \omega_0 $ is taken naively
with $ \Delta = - \langle \delta \omega \rangle $,
the effective squeezing $ \chi $ is significantly reduced,
even possibly becomes imaginary
with $ N_\gamma (t) \lesssim 1 $ oscillating
as $ \sin^2 | \chi | t $.

We have solved numerically the master equation
typically with $ \delta \omega (t)
= \langle \delta \omega \rangle ( 1 - \cos \Omega t ) $
and $ g(t) = - i \delta \omega (t) / 2 $
to confirm that the rotating-wave approximation is fairly good
for $ | \delta \omega (t) | \ll \omega_0 $.
The instantaneous-mode solution is also obtained
with $ \delta {\bar \omega} (t) = \delta \omega (t) $
and $ {\bar g} (t) = [ i / 2 {\bar \omega} (t) ] {\dot g} (t) $,
as seen so far.
It almost reproduces the rotating-wave approximation,
smoothing the actual small oscillation of $ N_\gamma (t) $
due to that of $ \delta \omega (t) $.
The relations  $ \delta {\bar \omega} (t) = \delta \omega (t) $
and $ {\bar g} (t) = [ i / 2 {\bar \omega} (t) ] {\dot g} (t) $
really imply $ | 2 \langle g \rangle_\Omega |
\simeq | 2 \langle {\bar g} \rangle_\Omega | $ (Fourier components)
around the resonance $ \Omega
= 2 ( \omega_0 + \langle \delta \omega \rangle ) $
for $ | \delta \omega (t) | \ll \omega_0 $,
giving essentially the same $ N_\gamma (t) $
in Eqs. (\ref{eqn:Ngamma}) and (\ref{eqn:chi}).

We now discuss the experimental realization
of DCE with the semiconductor plasma mirror.
It will be feasible with the sufficient maximal laser power
$ W_{\rm laser}^{\rm max} $
to achieve the enhanced displacement
as $ \delta_m^{\rm max}
\simeq [ ( n_e^{\rm max} e^2 / \epsilon_0 m_* ) / \omega_0^2 ] \delta
\sim 10^2 \delta $ or larger with $ \sin^2 kl = 1 $
(the slab placed in the middle of cavity $ l = L/2 $).
In this case, the conductivity effect $ \delta_m $
in Eq. (\ref{eqn:dlt-m})
dominates over the dielectric effect $ \delta_\epsilon $
in Eq. (\ref{eqn:dlt-e})
with $ \epsilon_1 (0) \sim 1 - 10 $
and $ \epsilon_1 (0) \leq | \epsilon_1 (t) | $
[even for the complex $ \epsilon_1 (t) $].
Then, we estimate roughly
$ \chi ( \Delta = 0 ) = | 2 \langle g \rangle_\Omega |
\sim \omega_0  ( \delta_m^{\rm max} / L ) \sim 10^{-2} \omega_0 $
for $ \delta \sim 10 \mu {\rm m} $ and $ L \sim 0.1 {\rm m} $.
This requires $ N_{\rm pulse} \gtrsim 100 $ repetitions
of laser pulse to create $ N_\gamma \gtrsim 10 $ photons
with $ \chi ( N_{\rm pulse} T ) \gtrsim 1 $.
The cavity $ Q $ value is reasonable as $ Q > \omega_0 / \chi \sim 10^2 $.
The tuning of $ \Omega $ for the resonance
should be made with the average shift
$ \langle \delta \omega \rangle \sim \chi \sim 10^{-2} \omega_0 $.
The time-profile of $ W_{\rm laser} (t) $ should also be chosen
suitably to optimize the Fourier component
$ \langle g \rangle_\Omega {\rm e}^{- i \Omega t} $ in $ g(t) $.
A detailed analysis will be made elsewhere
based on the present formulation.
The time-varying dielectric function $ \epsilon_1 (t) $ (complex)
and conductivity $ m_p^2 (t) $ are actually given
depending on the laser-power profile $ W_{\rm laser} (t) $.
By using these $ \epsilon_1 (t) $ and $ m_p^2 (t) $,
the frequency shift $ \delta \omega (t) $
and squeezing coupling $ g(t) $ are determined
in Eqs. (\ref{eqn:dom-pm}) and (\ref{eqn:g-pm}).
Then, the master equation is solved
to obtain the photon number $ N_\gamma (t) $.

{\it Detection}.--The photons created via DCE
can be detected suitably by Rydberg atoms
with principal quantum number $ n \approx 100 $
and transition frequency $ \sim {\rm GHz} $ \cite{D95-DK96,RITZ06}.
Rydberg atoms as two-level system are initially prepared
in the lower level, and injected into the cavity.
Some of these atoms are excited to the upper level
by absorbing the photons,
and detected outside the cavity as the signal of photons.
Recently, high-sensitivity measurement of blackbody radiation
has been performed at a frequency 2.527 GHz
and low temperatures 67 mK -- 1 K
by employing a Rydberg-atom cavity detector
with a newly developed selective field ionization scheme
for $ n \approx 100 $
(the atoms excited by absorbing photons
are selectively ionized by applying an electric field)
\cite{SPD06}.
It exceeds the standard quantum limit,
detecting less than one photon on average in the cavity.
Hence, the single-photon detection with Rydgerg atoms
is really capable of observing even a small number of DCE photons.
When $ N_{\rm Ryd} $ atoms are injected in the cavity,
the number of photons detected by atoms
is limited roughly as $ N_\gamma \lesssim N_{\rm Ryd} $
(acutally $ N_{\rm Ryd} \sim 100 $ \cite{SPD06}).
We also note that in order to observe purely the vacuum squeezing via DCE,
the cavity should be cooled well below 100 mK
to suppress the thermal photons as $ N_\gamma^{\rm thermal} \ll 1 $.

\begin{acknowledgments}
The authors appreciate valuable discussions
with S. Matsuki, Y. Kido, T. Nishimura, W. Naylor
and the Ritsumeikan University group.
This work was supported by KAKENHI (20340060).
\end{acknowledgments}

\end{document}